\documentclass[10pt,aps,prb,superscriptaddress,raggedbottom,showpacs,floatfix, twocolumn]{revtex4-1}

\usepackage{amsmath,amsfonts,amssymb,amsthm}
\usepackage[usenames,dvipsnames]{color}
\usepackage{graphicx,latexsym}

\newcommand{\ket}[1]{\left| #1 \right>} 

\newcommand{\Tsub}[1]{_{\mbox{\scriptsize #1}}}
\newcommand{\Tsup}[1]{^{\mbox{\scriptsize #1}}}

\begin{document}

\title{Rashba spin orbit interaction in a quantum wire superlattice}
\author{Gunnar Thorgilsson}
\affiliation{Science Institute, University of Iceland, Dunhaga 3, IS-107 Reykjavik, Iceland}
\author{J. Carlos Egues}
\affiliation{Departamento de F\'{i}sica e Inform\'{a}tica, Instituto de F\'{i}sica de S\~{a}o Carlos, Universidade de S\~{a}o Paulo, 13560-970 S\~{a}o Carlos, SP, Brazil}
\affiliation{Department of Physics, University of Basel, Klingelbergstrasse 82, CH-4056 Basel, Switzerland}
\author{Daniel Loss}
\affiliation{Department of Physics, University of Basel, Klingelbergstrasse 82, CH-4056 Basel, Switzerland}
\author{Sigurdur I. Erlingsson}
\affiliation{Reykjavik University, School of Science and Engineering, Menntavegi 1, IS-101 Reykjavik, Iceland}

\begin{abstract}
In this work we study the effects of a longitudinal periodic potential on a parabolic quantum
wire defined in a two-dimensional electron gas with Rashba spin-orbit interaction. For an infinite wire
superlattice we find, by direct diagonalization, that the energy gaps are shifted away from the usual Bragg
planes due to the Rashba spin-orbit interaction. Interestingly, our results show that the location of the band gaps in energy can be
controlled via the strength of the Rashba spin-orbit interaction. We have also calculated the charge conductance through a periodic
potential of a finite length via the non-equilibrium Green's function method combined with
the Landauer formalism. We find dips in the conductance that correspond well to the energy gaps
of the infinite wire superlattice. From the infinite wire energy dispersion, we derive an equation relating the location of the conductance
dips as a function of the (gate controllable) Fermi energy  to the Rashba spin-orbit coupling strength. We propose that the
strength of the Rashba spin-orbit interaction can be extracted via a charge conductance measurement.
\end{abstract}


\maketitle

\section{Introduction}
During the last two decades there has been much interest in using the electron spin in electronic
devices. This research field, often referred to as spintronics, has already made great impact on
metal-based information storage systems. There are hopes that a similar success can also be
achieved in semiconductor based systems \cite{Datta:apl1990,Awschalom:Book2002}.
Manipulating the spins of the electrons via external magnetic fields over nanometer length scales is
not considered feasible. Another, more attractive, method is to use electric fields to manipulate
electron spins via spin-orbit interaction.
The spin-orbit interaction arises from the fact that an electron moving in an external electrical
field experiences an effective magnetic field in its own reference frame, that in turn couples
to its spin via the Zeeman effect\cite{Greiner:Book1987}.

In condensed matter systems, the spin-orbit interaction is found in crystals with asymmetry in the underlying structure \cite{Winkler:book2010}.
In bulk this is seen in crystals without an inversion center (e.g zincblende structures) and is termed the Dresselhaus
spin-orbit interaction\cite{Dresselhaus:pr1955}. On the other hand the structural asymmetry of the confining potential in heterostructures gives 
rise to the so called Rashba term\cite{Bychkov:jpc1984}. The Rashba interaction has practical advantages in that it depends
on the electronic environment of the heterostructure which can be modified in sample fabrication and in-situ by gate voltages\cite{Engels:prb1997,Nitta:prl1997}.
This results in the possibility of varying the spin-orbit interaction on the nanometer scale. Interestingly, even structurally symmetric heterostructures can
present spin-orbit interaction provided that coupling between subbands of distinct parities is allowed\cite{Bernarder:prb2007,Calsaverini:prb2008}.

The spin-orbit strength can be measured in a variety of different experimental setups\cite{Zawadzki:SST2004}:
In a magnetoresistance measurement via Shubnikov-de Haas
oscillations\cite{Engels:prb1997,Nitta:prl1997,Matsuyama:prb2000,Guzenko:prb2007,Simmonds:jap2008}, 
weak (anti-) localization \cite{Koga:prl2002,Miller:prl2003,Guzenko:prb2007,Yu:prb2008}, or electron spin resonances
in semiconducting nanostructures\cite{Kato:nature2004,Duckheim:naturephysics2006,Meier:naturephysics2007,Duckheim:prb2009} or quantum dots\cite{Golovach:prb2006},
or optically via spin relaxation \cite{Eldrige:prb2008}, spin precession \cite{Kato:nature2004},
spin-flip Raman scattering \cite{Jusserand:prl1992}, or radiation-induced magnetoresistance oscillations \cite{Mani:prb2004}.

\begin{figure}[htp]
 \centering
\includegraphics[width=0.33\textwidth]{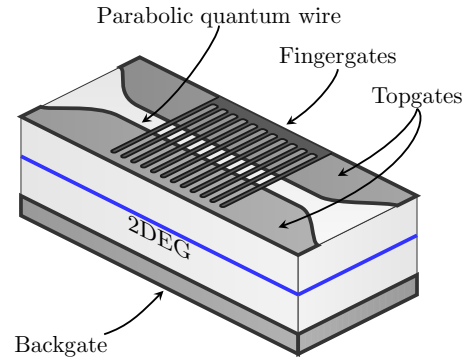}
 \caption{Schematic of the proposed experimental setup. The top gates produce the parabolic confinement
         of the quantum wire while the finger gates produce the longitudinal periodic potential. The Rashba spin-orbit interaction 
         is controlled by the backgate.}
\label{fig:SchematicPic}
\end{figure}

In this paper we propose a method for extracting the strength of the Rashba spin-orbit interaction via a charge conductance measurement. This method
does not require the use of external magnetic fields or radiation sources. We consider a  quantum wire modulated by an external periodic potential,
Fig. \ref{fig:SchematicPic}. Essentially, the method relies on the fact that the Rashba-induced shifts of the band gap positions in energy dramatically
alter the charge conductance of the superlattice.

Via direct diagonalization, we  determine the band structure of an infinite parabolically confined quantum wire. 
The energy bands clearly show band gaps that are renormalized by the Rashba interaction.
Interestingly, the band gaps shift in energy as the strength of the Rashba interaction is varied.
The location of the band gaps are
at the crossing of energy bands from adjacent Brillouin zones. These energy bands can be calculated via an analytical approximation 
scheme\cite{Erlingsson:prb2010}.

Using Non-Equilibrium Green's Functions (NEGF) and the Landauer formalism we calculate the charge conductance through a finite region containing both a periodic potential and the Rashba interaction.
In the conductance we find several dips appearing at different location in energy. 
Moreover, these conductance dips coincide with renormalized band gaps of the superlattice.
As with the band gaps the positions of the conductance dips are at the crossing points of the energy bands from next neighbor Brillouin zones.
For a wide range of Rashba spin-orbit interaction strengths some of the conductance dips shift linearly in energy as function of the strength of the Rashba coupling.
For this range we derive a relation, Eq.\ (\ref{eq:linearEalpha}), describing the location of linearly shifting dips in energy
as a function of Rashba interaction strength. The Rashba coupling can therefore be extracted by fitting the
shift of conductance dips via Eq.\ (\ref{eq:linearEalpha}). Figure 1 shows a schematic of the proposed experimental setup.

This paper is organized as follows. In Sec.\ \ref{sec:InfWire} we calculate the energy bands of the infinite wire superlattice via direct diagonalization.
By using an analytical
approximation scheme\cite{Erlingsson:prb2010} we then show how the resulting band gaps depend on the Rashba spin-orbit interaction.
Via the NEGF method and the Landauer formalism we introduce in Sec.\ \ref{sec:FinWire} a numerical scheme to calculate
the charge conductance through a finite region. This region is connected to electron reservoirs to the left and right,
and contains both a periodic potential and a Rashba spin-orbit coupling. 
In Sec.\ \ref{sec:Result} we then show that the band picture of the infinite wire superlattice, developed in Sec.\ \ref{sec:InfWire},
is applicable to the finite length periodic potential. We close Sec.\ \ref{sec:Result} with a discussion about possible experimental procedures.
Lastly, in Sec. \ref{sec:PerVar} we show that our results are robust against fluctuations in the strength and width of the periodic potential. 

\section{Model system: The infinite wire superlattice}
\label{sec:InfWire}
We investigate an infinite quasi-1D parabolic wire with a uniform Rashba spin-orbit coupling
in the presence of a longitudinal modulation described by the potential
\begin{equation}
V\Tsub{p}(x)=V\Tsub{p0}\sum_n c_n e^{in\frac{2\pi}{\lambda}x},
\label{eq:PeriodicEq}
\end{equation}
where $\lambda$ is the period of the superlattice. The Hamiltonian that describes this system is
\begin{align}
H&=\frac{1}{2m^*}(p_x^2+p_y^2)+\frac{1}{2}m^*\omega_0^2y^2\nonumber\\
 &+\frac{\alpha}{\hbar}(p_y\hat{\sigma}_x-p_x\hat{\sigma}_y)+V\Tsub{p}(x),
 \label{eq:originalH}
\end{align}
were $m^*$ is the effective mass, $p_x$ and $p_y$ are the momentum operators in the longitudinal and
transverse direction of the wire, $\alpha$ is the Rashba spin-orbit strength, and $\omega_0$ is the 
confinement frequency of the parabolic potential.

To find the eigenvalues of the Hamiltonian in Eq.\ (\ref{eq:originalH}) it is convenient to introduce 
the standard ladder operator $\hat{a}$ of the parabolic confinement and rotate the
spin operators so that the $p_x$ part in the Rashba interaction term couples to the $\hat{\sigma}_z$ operator\cite{Erlingsson:prb2010} 
\begin{align}
\widetilde{H}&=e^{-i\pi\hat{\sigma}_x/4}He^{i\pi\hat{\sigma}_x/4}\nonumber\\
         &=\frac{1}{2}\left(k-\frac{2\pi}{\lambda}n\right)^2+\frac{1}{2}+\hat{a}^{\dagger}\hat{a}
          -q\Tsub{R}\left(k-\frac{2\pi}{\lambda}n\right)\hat{\sigma}_z\nonumber\\
         &+\frac{iq\Tsub{R}}{\sqrt{8}}[\hat{\sigma}_+(\hat{a}^{\dagger}-\hat{a})-\mathrm{h.c.}]
          +V\Tsub{p}(x)\nonumber\\
         &=\widetilde{H}_0+\widetilde{H}_1,
\label{eq:rotatedH}
\end{align}
here $\hat{\sigma}_+=\hat{\sigma}_x+i\hat{\sigma}_y$ is the spin ladder operator,
$q\Tsub{R}=\frac{m^*\alpha}{\hbar^2}$ is the rescaled Rashba strength and $\hbar \left(k-\frac{2\pi}{\lambda}n\right)$
are the eigenvalues of the operator $p_x$. In Eq.\ (\ref{eq:rotatedH}) we scale all lengths
in oscillator length $l=\sqrt{\hbar/m^*\omega_0}$ and all energies in $\hbar\omega_0$.
We separate the Hamiltonian $\widetilde{H}$ into a diagonal part,
\begin{align}
\widetilde{H}_0&=\frac{1}{2}\left(k-\frac{2\pi}{\lambda}n\right)^2
               +\frac{1}{2}+\hat{a}^{\dagger}\hat{a}-q\Tsub{R}\left(k-\frac{2\pi}{\lambda}n\right)\hat{\sigma}_z\nonumber\\
              &+c_0V\Tsub{p0},
\end{align}
and a non-diagonal part 
\begin{equation}
\widetilde{H}_1=\frac{iq\Tsub{R}}{\sqrt{8}}[\hat{\sigma}_+(\hat{a}^{\dagger}-\hat{a})-\mathrm{h.c.}]
          +\sum_{n\neq0} c_n e^{in\frac{2\pi}{\lambda}}.
\end{equation}
%
%

\subsection{Zeroth-order eigenstates and eigenvalues}
\label{sec:H0}
The eigenstates of the $\widetilde{H}_0$ Hamiltonian are represented by the kets $\ket{k,m,s}$.
Here $m$ is the quantum number of the harmonic transverse energy bands, i.e. the eigenvalue of
the $\hat{a}^{\dagger}\hat{a}$ operator, and $s$ is the eigenvalue of the $\hat{\sigma}_z$ operator
with $s=+1$ and $s=-1$ denoting the spin up and spin down states, respectively. The corresponding
eigenenergies are 
\begin{align}
 E^n_{m,s}(k)&=\frac{1}{2}\left(k-\frac{2\pi}{\lambda}n\right)^2+\frac{1}{2}+m-sq\Tsub{R}\left(k-\frac{2\pi}{\lambda}n\right)\nonumber\\
             &+c_0V\Tsub{p0}.
\end{align}

\begin{figure}[htp]
 \centering
\includegraphics[width=0.49\textwidth]{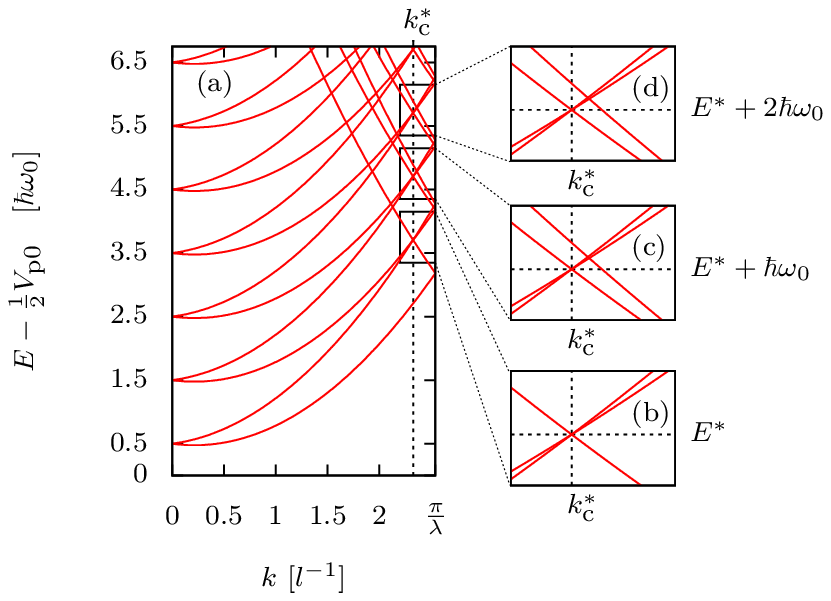}
\caption{(a) Energy bands of the zeroth-order Hamiltonian $\widetilde{H}_0$ for $l=50$ nm,
$\lambda=62$ nm$=1.2 l$, $\alpha=8$ meVnm ($q\Tsub{R}=0.21 l^{-1}$), $k\Tsub{c}^*=2.3l^{-1}$, $c_0=1/2$, and $V\Tsub{p0}=0.5$.
(b) Crossing of the energy bands corresponding to the $\ket{k,0,-1}$, $\ket{k,1,+1}$,
and $\ket{k-2\pi/\lambda,0,-1}$ states. (c) Crossing of the energy bands corresponding to the
$\ket{k,1,-1}$, $\ket{k,2,+1}$, $\ket{k-2\pi/\lambda,1,-1}$, and $\ket{k-2\pi/\lambda,0,+1}$ states.
(d) Same as (c) but for higher bands, i.e. $m\rightarrow m+1$.}
\label{fig:H0EnBands}
\end{figure}

A plot of the $E^n_{m,s}(k)$ energy bands vs $k$ in half of the Brillouin zone, i.e., $k=0\ldots\pi/\lambda$, can be seen in Fig.\ \ref{fig:H0EnBands} (a).

The Hamiltonian $\widetilde{H}_1$ introduces couplings \emph{i)} between the $\ket{k,m,s}$ and
the $\ket{k,m\pm1,\overline{s}}$ states, due to the Rashba interaction, and \emph{ii)} between 
the $\ket{k,m,s}$ and the $\ket{k\pm n2\pi/\lambda,m,s}$ states, due to the periodic potential. Note that the coupling is strongest
where the energy bands corresponding to these states cross each other. For some particular Rashba strength $\alpha^*$ 
the energy bands associated with the states $\ket{k,0,-1}$ and $\ket{k,1,+1}$ of
$\widetilde{H}_0$ cross at $k\Tsub{c}^*=1/(2q\Tsub{R}^*)$, where $q\Tsub{R}^*=\alpha^*m^*/\hbar^2$ .
If we choose the period $\lambda$ of the superlattice potential, Eq.\ (\ref{eq:PeriodicEq}), as
\begin{equation}
\lambda=2\pi\frac{q\Tsub{R}^*}{1+2(q\Tsub{R}^*)^2},
\label{eq:period}
\end{equation}
the $\ket{k-2\pi/\lambda,0,-1}$ energy band will also cross at $k\Tsub{c}^*$. This crossing point occurs in energy at
\begin{equation}
E^*=\frac{1}{8q\Tsub{R}^*}+1+c_0V\Tsub{p0},
\end{equation}
see Fig.\ \ref{fig:H0EnBands} (b).
In the following we refer to this particular choice of parameter $\alpha^*$ as the reference spin-orbit coupling strength.
Similar crossing also occur at $k\Tsub{c}^*$ with the energy bands associated with the states $\ket{k,m,-1}$,
$\ket{k,m+1,+1}$, and $\ket{k-2\pi/\lambda,m,-1}$ for energies $E^*+m\hbar\omega_0$ with the
energy band corresponding to the state $\ket{k-2\pi/\lambda,m-1,+1}$ crossing close by. The crossings for $m=1$ and $m=2$
can be seen, respectively, in Fig.\ \ref{fig:H0EnBands} (c) and (d).

\subsection{Coupling between eigenstates}
\begin{figure}[htp]
 \centering
 \includegraphics[width=0.47\textwidth]{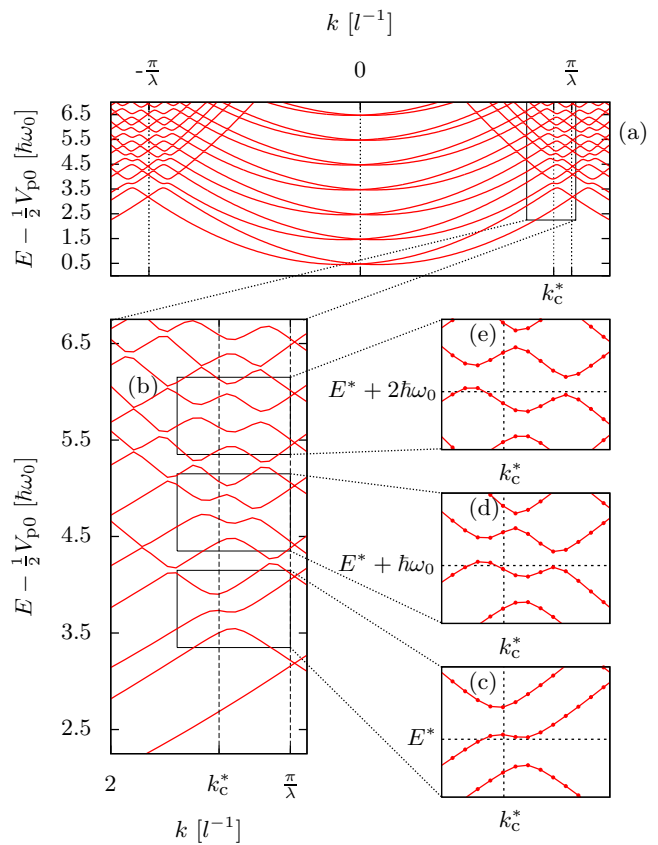}
\caption{(a) Energy bands of the Hamiltonian $\widetilde{H}= \widetilde{H}_0+ \widetilde{H}_1$
for the same parameters as used in Fig.\ \ref{fig:H0EnBands}.
The energy bands are calculated via direct diagonalization of $\widetilde{H}$.
(b) A blow up focusing on the band gaps of the first three crossings at $k\Tsub{c}^*$.
(c) The band gaps formed at the energy band crossing shown in Fig.\ \ref{fig:H0EnBands} (b). 
In (d) and (e), the energy band crossings shown in Fig.\ \ref{fig:H0EnBands} (c) and (d), respectively.}
\label{fig:HFullEnBands}
\end{figure}

We calculate the eigenenergies of the full Hamiltonian, $\widetilde{H}$, via direct diagonalization. The
resulting energy bands are plotted in Fig.\ \ref{fig:HFullEnBands} (a).
The energy bands that cross at $k\Tsub{c}^*$ and the one crossing close by, see Sec.\ \ref{sec:H0}, corresponds
to the states coupled by $\widetilde{H}_1$. A blow up of the resulting energy gaps can be seen
in Fig.\ \ref{fig:HFullEnBands} (b). At $E^*$ the coupling results in a double energy gap, 
see Fig.\ \ref{fig:HFullEnBands} (c), and a triple energy gap at the higher energy crossings, 
see Fig.\ \ref{fig:HFullEnBands} (d) and (e). In Fig.\ \ref{fig:HFullEnBands} the spin-orbit
strength is at the reference value $\alpha^*$.

Note that for a non-zero Rashba coupling, as in Fig.\ \ref{fig:HFullEnBands}, the energy gaps have shifted from the
Bragg plane at $k=\pi/\lambda$. This is because the spin-orbit interaction shifts the wave-number $k$ of the
electrons in the longitudinal direction and thus renormalizes the locations of interferences.

When the strength of the Rashba interaction is changed the crossing points $k\Tsub{c}$ of the energy-bands shift
and with them the energy gaps. These crossing points can be worked out analytically by using the eigenenergies
of the effective Rashba Hamiltonian of the parabolic wire\cite{Erlingsson:prb2010}, which are
\begin{align}
\varepsilon^n_{m,\uparrow}(k,q\Tsub{R})=&\frac{\left(k-n\frac{2\pi}{\lambda}\right)^2}{2}+m
                             -\frac{q\Tsub{R}^2}{2(1+2q\Tsub{R}\left(k-n\frac{2\pi}{\lambda}\right))}\nonumber\\
                             &+\Delta_m\left(k-n\frac{2\pi}{\lambda},q\Tsub{R}\right)+c_0V\Tsub{p0},
\label{eq:EffEnUp}
\end{align}
and 
\begin{align}
\varepsilon^n_{m,\downarrow}(k,q\Tsub{R})=&\frac{\left(k-n\frac{2\pi}{\lambda}\right)^2}{2}+m+1
                                -\frac{q\Tsub{R}^2}{2(1+2q\Tsub{R}\left(k-n\frac{2\pi}{\lambda}\right))}\nonumber\\
                               &-\Delta_{m+1}\left(k-n\frac{2\pi}{\lambda},q\Tsub{R}\right)+c_0V\Tsub{p0},
\label{eq:EffEnDown}
\end{align}
for $k-2\pi/\lambda\geq0$. For $k-2\pi/\lambda <0$  $\varepsilon_{m,s}(k,q\Tsub{R})=\varepsilon_{m,-s}(-|k|,q\Tsub{R})$.
In Eq.\ (\ref{eq:EffEnUp}) and (\ref{eq:EffEnDown})
\begin{align}
 \Delta_m(k,q\Tsub{R})=&\frac{1}{2}\left[\left(1-2q\Tsub{R}k-\frac{q\Tsub{R}^2m}{1+2q\Tsub{R}k}\right)^2\right. \nonumber\\
             &+\left.2q\Tsub{R}^2m\left(1-\frac{q\Tsub{R}^2m}{4(1+2q\Tsub{R}k)}\right)^2\right]^{1/2},
\end{align}
and we have added the energy constant $c_0V\Tsub{p0}$ resulting from the periodic potential and replaced $k$ with $k-n2\pi/\lambda$.
Note that the energy bands described by Eq.\ (\ref{eq:EffEnUp}) and Eq.\ (\ref{eq:EffEnDown}) are derived for wires without a 
longitudinal periodic potential and therefore do not contain energy gaps that result from the periodic potential, i.e. they are
the solution of the $\widetilde{H}$, see Eq.\ (\ref{eq:rotatedH}), with $c_0\neq0$ and $c_n=0$ for $n\neq0$. Having determined the crossing points,
we insert them into either of the crossing energy bands, Eq.\ (\ref{eq:EffEnUp}) or Eq.\ (\ref{eq:EffEnDown}),
to obtain the location of the crossing points in energy as a function of $\alpha$. 
In the appendix we present the equation for the crossing point between the energy bands $\varepsilon^0_{m,\downarrow}(k,q\Tsub{R})$ and $\varepsilon^1_{m-1,\uparrow}(k,q\Tsub{R})$
corresponding to states $\ket{k,m+1,+1}$ and $\ket{k-2\pi/\lambda,m,-1}$, see Eq.\ (\ref{eq:kCross}).
To determine the location in energy of this crossing point as a function of $\alpha$ we insert $k\Tsub{c}$ from Eq.\ (\ref{eq:kCross})
into either $\varepsilon^0_{m,\downarrow}(k,q\Tsub{R})$ or $\varepsilon^1_{m-1,\uparrow}(k,q\Tsub{R})$.
\begin{figure}[htp]
\centering
 \includegraphics[width=0.45\textwidth]{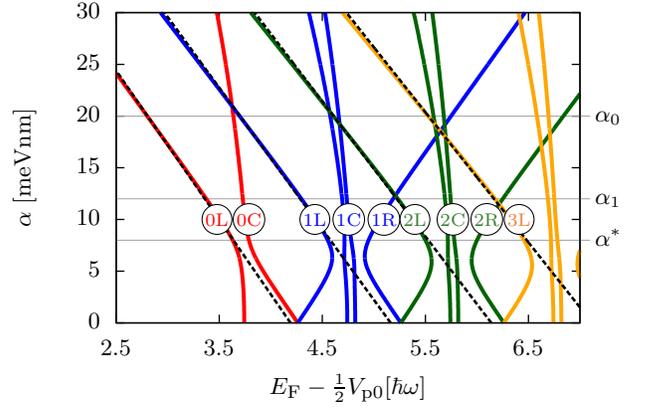}
\caption{(Color online) Trajectories of the crossing points between the energy bands.
The black dashed lines are the linearization of the left moving trajectories around $\alpha_0$. 
We also mark the values $\alpha^*=8$ meVnm and $\alpha_1=12$ meVnm onto the $y$-axis. These values are used for the conductance results
in Fig.\ \ref{fig:gapVsG}.}
\label{fig:Trajectories}
\end{figure}

Now, as the crossing points shift in energy with $\alpha$ they can be thought of as trajectories.
The trajectories of the crossing points, calculated from Eq.\ (\ref{eq:EffEnUp}) and (\ref{eq:EffEnDown}), can be seen in
Fig.\ \ref{fig:Trajectories}. For higher energies there are further trajectories.

We index the trajectories via $m$ and the letters ``C'', ``L'', and ``R''. The trajectories that we label with the letter C
only shift a little to the left for $\alpha > \alpha^*$ (see lowest horizontal line in Fig.\ \ref{fig:Trajectories}).
Relative to the C labeled trajectories and again for $\alpha > \alpha^*$, we see trajectories that make a large shift to the left and right.
Those trajectories we, respectively, label with the letters ``L'' and ``R''.

For spin-orbit strengths close to the reference strength, $\alpha^*$, the crossing points follow
nonlinear trajectories. As the spin-orbit coupling becomes larger or lower than the reference strength
the trajectories quickly become more linear.
The choice of the reference Rashba spin-orbit strength $\alpha^*$ determines the period of the periodic potential,
see Eq.\ (\ref{eq:period}).

\subsection{Extracting the Rashba coupling}
We will show in Sec.\ \ref{sec:FinWire} that the band gaps appear as dips in the charge conductance through a finite periodic potential.
By fitting the measured energy shift of the conductance dips to the trajectories in Fig.\ \ref{fig:Trajectories} it is possible to extract the value
of $\alpha$. The linear parts of the trajectories are best suited for fitting. It would therefore be convenient that the range of $\alpha$
extracted via the fitting is contained within the linear region. This can be achieved by choosing a sufficiently low $\alpha^*$-value
(and thus $\lambda$, Eq.\ \ref{eq:period}).
We present below a linearized equation for the $\alpha$-value of the crossing point between the
energy bands corresponding to the states $\ket{k,m+1,+1}$ and $\ket{k-2\pi/\lambda,m,-1}$ as a function of Fermi energy.
By Taylor expanding around some point $\alpha_0$ in the linear region of the trajectories and making a linear approximation 
we obtain
\begin{equation}
\alpha=\frac{\hbar^2}{m^*}\frac{\hbar}{l}\left[
q\Tsub{R0}
-\frac{E\Tsub{F}-\frac{1}{2}V\Tsub{p0}- F(m)}
      { G(m)}\right],
\label{eq:linearEalpha}
\end{equation}
here $F(m)$ and $G(m)$ are known functions of the energy band index $m$, see Eq.\ (\ref{eq:F}) and Eq.\ (\ref{eq:G}).
The derivation of Eq.\ (\ref{eq:linearEalpha}) is shown in the appendix. In Fig.\ \ref{fig:Trajectories} we plot as dashed lines
linearized trajectories described by Eq.\ (\ref{eq:linearEalpha}) where we have chosen $\alpha_0=20$ meVnm.

\section{Finite periodic potential}
\label{sec:FinWire}
In this section we calculate in the linear response the conductance through a finite periodic potential.
This is done via the Landauer formula together with the NEGF method.

\subsection{The system setup}
\begin{figure}[htp]
 \centering
 \includegraphics[width=0.47\textwidth]{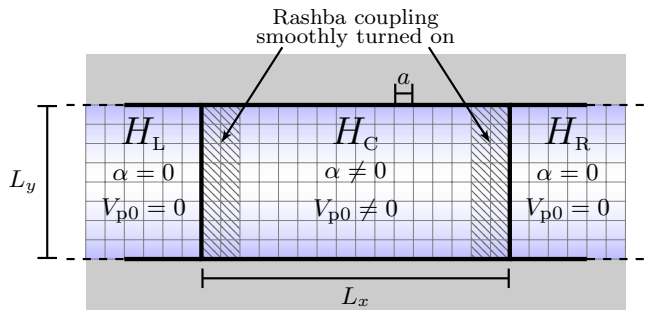}
\caption{Schematic of the system used in our numerical simulations. The system is divided into a central region of length $L_x$, described by
the Hamiltonian $H_C$ and two semi-infinite wires, described by the Hamiltonians $H\Tsub{L}$ and $H\Tsub{R}$.
The total system is discretized on a grid with mesh size $a$. At the left and right edges of the central region the 
Rashba spin-orbit coupling is smoothly turned on.}
\label{fig:NumSchema}
\end{figure}

We consider a hardwalled wire of width $L_y$ with a transverse parabolic potential. We divide
the wire into a finite central region of length $L_x$ and semi-infinite left and right parts,
see Fig.\ \ref{fig:NumSchema}.
The central region includes a Rashba spin-orbit interaction described by a symmetrized
Hamiltonian which is turned on smoothly, at both the left and right ends.
In the central region we also assume a longitudinal periodic potential that represents the potential due to the fingergates.
We describe the central region by the Hamiltonian
\begin{align}
H\Tsub{C}&=\frac{1}{2m^*}(p_x^2+p_y^2)+\frac{1}{2}m^*\omega_0^2y^2\nonumber\\
 &+\frac{1}{2\hbar}\left\lbrace\alpha(x,y),p_y\hat{\sigma}_x-p_x\hat{\sigma}_y\right\rbrace\nonumber\\
 &+V\Tsub{p0}\frac{1}{2}\left(1-\cos\left(\frac{2\pi}{\lambda}x\right)\right)
\end{align}
Here $\{,\}$ denotes an anticommutator.

The left and right leads contain the same parabolic potential as in the center region
but neither a Rashba spin-orbit interaction nor a periodic potential in the longitudinal direction
, i.e., they are described by the Hamiltonian
\begin{equation}
H_{\mathrm{L/R}}=\frac{1}{2m^*}(p_x^2+p_y^2)+\frac{1}{2}m^*\omega_0^2y^2.
\end{equation}

The total system, $H\Tsub{T}=H\Tsub{L}+H\Tsub{C}+H\Tsub{R}$, is discretized via the finite difference method on
a grid of $N_x\times N_y$ points with a mesh size $a$. A schematic of the system can be seen in Fig.\ \ref{fig:NumSchema}.

\subsection{The numerical formalism}
Here we use the NEGF method\cite{Datta:book1995,Ferry:book1997,Pastawski:pmf2001} to calculate the charge conductance.
The method requires us to find the retarded Green's function of the central region $\mathbf{G}\Tsub{C}\Tsup{r}$.
To do this we have to isolate $\mathbf{G}\Tsub{C}\Tsup{r}$ from the infinite matrix equation describing the retarded Green's function
of the total system
\begin{equation}
(E\mathbf{I}-\mathbf{H}_\mathrm{T})\mathbf{G}\Tsub{T}\Tsup{r}=\mathbf{I}.
\label{eq:InfiniteGreen}
\end{equation}
By separating the total Green's function into a left, right, and central part we can write Eq.\ (\ref{eq:InfiniteGreen}) as
\begin{small}
\begin{align}
&\left[\begin{array}{ccc}
                 E\mathbf{I}-\mathbf{H}\Tsub{L} & -\mathbf{H}\Tsub{LC}  &    0     \\ 
                  -\mathbf{H}\Tsub{CL} & E\mathbf{I}-\mathbf{H}\Tsub{C} & -\mathbf{H}\Tsub{CR}  \\
                      0   & -\mathbf{H}\Tsub{RC}  & E\mathbf{I}-\mathbf{H}\Tsub{R} \\
                \end{array}\right]
          \left[\begin{array}{lll}
                 \mathbf{G}\Tsub{L}\Tsup{r} & \mathbf{G}\Tsub{LC}\Tsup{r} & \mathbf{G}\Tsub{LR}\Tsup{r} \\ 
                 \mathbf{G}\Tsub{CL}\Tsup{r} & \mathbf{G}\Tsub{C}\Tsup{r} & \mathbf{G}\Tsub{CR}\Tsup{r} \\
                 \mathbf{G}\Tsub{RL}\Tsup{r} & \mathbf{G}\Tsub{RC}\Tsup{r} & \mathbf{G}\Tsub{R}\Tsup{r} \\
                \end{array}\right]\nonumber\\
        =&\left[\begin{array}{ccc}
                 \mathbf{I} & 0 & 0 \\ 
                 0 & \mathbf{I} & 0 \\
                 0 & 0 & \mathbf{I} \\
                \end{array}\right].
\label{eq:HGmatrixEq}
\end{align}
\end{small}
The matrices $\mathbf{H}\Tsub{Cj}=\mathbf{H}\Tsub{jC}=t\mathbf{I}$, couple together
the central region to the left ($j=$ L) and right leads ($j=$ R). Here $t=\hbar/2m^*a^2$ is the tight-binding
hopping parameter that results from the discretization.
Multiplying out Eq.\ (\ref{eq:HGmatrixEq}) gives nine matrix equations
from which we can isolate a \emph{finite} matrix equation for $\mathbf{G}\Tsub{C}\Tsup{r}$.
This is done by treating the contributions from the infinite leads as 
self-energies\cite{Datta:book1995,Ferry:book1997,Pastawski:pmf2001,LopezSancho:jpf1985}.
The matrix equation for $\mathbf{G}\Tsub{C}\Tsup{r}$ is
\begin{equation}
 \big(E\mathbf{I}-\mathbf{H}\Tsub{C}-\sum_j\mathbf{\Sigma}_j\big)\mathbf{G}\Tsup{r}\Tsub{C}
 =\mathbf{I},
 \label{eq:FiniteGreen}
\end{equation}
and the self-energy of lead $j=$L, R is $\mathbf{\Sigma}_j=\mathbf{H}\Tsub{C$j$}\mathbf{G}_{jj}\mathbf{H}\Tsub{$j$C}$ where $\mathbf{G}_{jj}$ is 
the Green's function of lead $j$ and is determined analytically\cite{Datta:book1995}.
Here $\mathbf{H\Tsub{C}}$ is the discretized Hamiltonian of Eq.\ (\ref{eq:originalH}) over the central region.
All the matrices are $2N_xN_y \times 2N_xN_y$ matrices. In order to save computational power only the necessary
Green's function matrix elements are calculated with the recursive Green's function method\cite{Ferry:book1997}.
Here we are interested in low biases  and hence focus on the linear response regime. From the Green's function
the charge conductance at the Fermi Energy $E\Tsub{F}$ can be calculated, via 
the Fisher-Lee relation\cite{Fisher:prb1981}, as
\begin{equation}
G(E\Tsub{F})=\frac{2e^2}{h}
\mbox{Tr}\left[\boldsymbol\Gamma\Tsub{R} \mathbf{G}\Tsup{r}\Tsub{C}\boldsymbol\Gamma\Tsub{L}\mathbf{G}\Tsup{a}\Tsub{C}\right],
\label{eq:Conductance}
\end{equation}
where  $\boldsymbol\Gamma\Tsub{j}=-2\mbox{Im}\left(\boldsymbol\Sigma\Tsub{j}\right)$
and $\mathbf{G}\Tsup{a}\Tsub{C}=\left(\mathbf{G}\Tsup{r}\Tsub{C}\right)^{\dagger}$.
In what follows we use Eq.\ (\ref{eq:Conductance})  to calculate the charge conductance through our system.

\subsection{Numerical values used in the calculation}
In our simulations we consider a Ga$_{1-x}$In$_x$As alloy with an effective mass $m^*=0.041m\Tsub{e}$,
with $m_e$ being the bare electron mass.
The width of the wire is set as $L_y=500$ nm and its length as $L_x=4.92\ \mu$m. The oscillator length is $l=50$ nm which
corresponds to $\hbar\omega_0=0.74$ meV.
For the discretization we use $N_x\times N_y=600\times63$ points with $a=8.1$ nm being the distance between nearest neighbors.
This corresponds to a tight-binding hopping parameter $t=29$ meV$=39\ \hbar\omega_0$.  We choose the reference Rashba strength
as $\alpha^*=8$ meV nm (i.e. $k\Tsub{R}^*= 0.21\ l^{-1}$) which results in an energy band crossing point at
$k\Tsub{c}^*=1/(2q\Tsub{R}^*)=2.3\ l^{-1}$ and a period
$\lambda=\pi/\left(k\Tsub{c}^*+q\Tsub{R}\right)= 1.23\ l = 7.62\ a=61.5$ nm. This corresponds to $L_x/\lambda=$80 fingergates.
The strength of the periodic potential is set as $V\Tsub{p0}=0.5\hbar\omega_0=0.37$ meV.

\section{Results}
\label{sec:Result}
\begin{figure}[htp]
\centering
\includegraphics[width=0.45\textwidth]{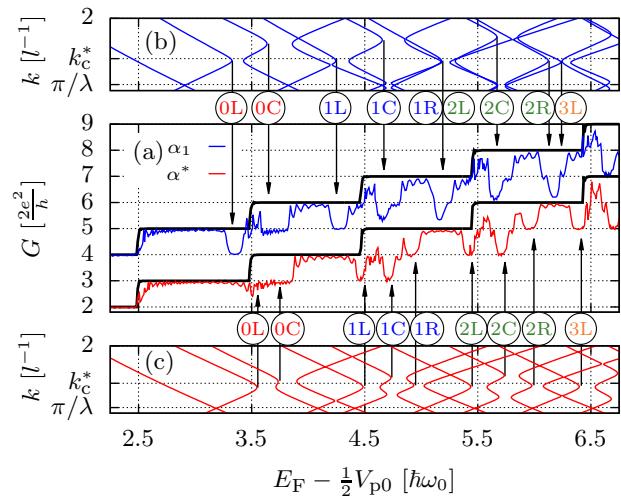}
\caption{(Color online) Comparison between energy gaps of the infinite superlattice and conductance dips of the finite periodic potential.
In (a) the results from the numerical calculation of the conductance is shown for $\alpha_1=12$ meVnm, and $\alpha^*=8$ meVnm.
Note that the conductance curve for $\alpha_1$ has been shifted by 2$\times2e^2/h$ for clarity.
The results for a wire with a non-periodic potential $V\Tsub{p}(x)=1/2V\Tsub{p0}$ is plotted with
black solid lines.
In (b) and (c) the energy bands of the infinite wire superlattice are plotted for the same $\alpha$ values as in (a), respectively.
Note that (c) is the same figure as Fig.\ \ref{fig:HFullEnBands} (b), but rotated clockwise by $90^{\circ}$.
The correspondence between the nine pairs of dips and band gaps are indicated by the labeled arrows. 
}
\label{fig:gapVsG}
\end{figure}

In Fig.\ \ref{fig:gapVsG} we compare the calculated conductance,  Fig.\ \ref{fig:gapVsG} (a), through the finite wire to the energy
bands of the infinite wire, Fig.\ \ref{fig:gapVsG} (b) and (c), as a function of Fermi energy. Results for two values of $\alpha$ 
are presented, $\alpha^*=8$ meVnm and $\alpha_1=12$ meVnm.
For comparison we also plot in  Fig.\ \ref{fig:gapVsG} (a) the conductance through a non-periodic potential
$V\Tsub{p}(x)=1/2V\Tsub{p0}$ for the same values $\alpha$ mentioned above, i.e. the corresponding gapless systems\cite{Mireles:prb2001}.
Each energy gap is associated to its corresponding conductance dip via a labeled arrow. 
The labeling on the arrows is the same labeling as those of the trajectories in Fig.\ \ref{fig:Trajectories}.
We see that there is a good correspondence between the dips in conductance and the gaps in the energy band for both $\alpha^*$ and $\alpha_1$.
This indicates that the behavior of the \emph{finite} periodic potential can be properly described with the band model from the infinite periodic potential.
In Fig.\ \ref{fig:Trajectories} we can see how the band gaps shift in energy as a function of $\alpha$. 
We want see if the energy shift of conductance dips correlates with the shift of the band gaps.
This is most easily seen by considering the differential conductance 
\begin{equation}
\frac{\partial G}{\partial E\Tsub{F}}\approx\frac{G(E\Tsub{F$_{i+1}$})-G(E\Tsub{F$_i$})}{E\Tsub{F$_{i+1}$}-E\Tsub{F$_i$}}.
\end{equation}

\begin{figure}[htp]
  \centering
  \includegraphics[width=0.47\textwidth]{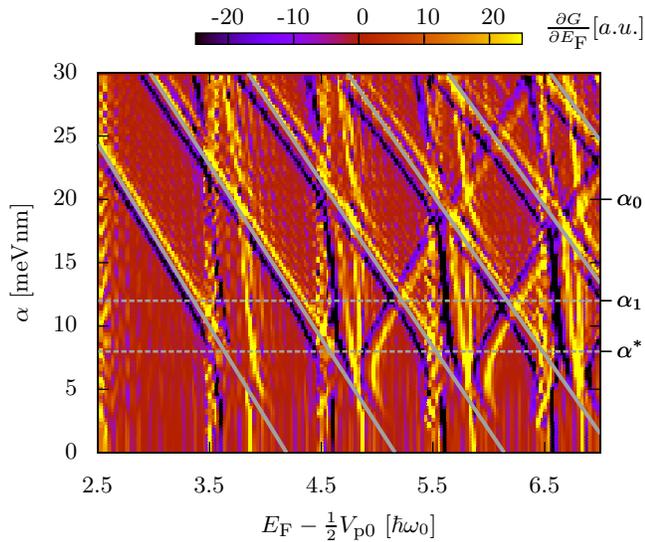}
  \caption{(Color online) Plot of the differential conductance $\partial G/\partial E\Tsub{F}$ through the finite superlattice as a function
           of the Fermi energy $E\Tsub{F}$ and spin-orbit coupling, $\alpha$. The linearized trajectories of the crosspoints between
           the energy bands corresponding to the $\ket{k,m+1,+1}$ and $\ket{k-2\pi/\lambda,m-1,+1}$ states are shown
           with gray solid lines. The trajectories are linearized around $\alpha_0=20$ meVnm. The Rashba spin-orbit values,
           $\alpha^*=8$ meVnm and $\alpha_1=12$ meVnm, corresponding to the curves in Fig.\ \ref{fig:gapVsG} are marked on the right border.
           Note that $q\Tsub{R}=m^*\alpha/\hbar^2l$ and that the differential conductance $\partial G / \partial E\Tsub{F}$ is in arbitrary units.}
 \label{fig:StandardSurface}
 \end{figure}

In Fig.\ \ref{fig:StandardSurface} we plot $\frac{\partial G}{\partial E\Tsub{F}}$ as a function of $E_F$ and $\alpha$.
There we see the conductance steps of the parabolic wire as relatively straight vertical trajectories
appearing at the tic marks on the horizontal axis.
The rest of the trajectories in Fig.\ \ref{fig:StandardSurface} correspond to dips in the charge conductance.
By comparing the trajectories, that the dips in charge conductance form, to the ones of the band gaps in Fig.\ \ref{fig:Trajectories}
we see an excellent match.
This firmly confirms that the conductance dips of the finite wire can be described via the band model for the superlattice.
We also plot in Fig.\ \ref{fig:StandardSurface} as light-gray lines the linearized trajectories of crossing points between the
energy bands corresponding to the $\ket{k,m+1,+1}$ and $\ket{k-2\pi/\lambda,m,-1}$ states, see Eq.\ (\ref{eq:linearEalpha}).
The trajectories are linearized around $\alpha$=20 meVnm and are identical to the straight dashed lines in Fig. \ref{fig:Trajectories}.

\subsection{Discussion of possible experimental procedures}
\label{sec:Experomento}
From Fig.\ \ref{fig:StandardSurface} we can extract the Rashba spin-orbit coupling by fitting the linear 
energy shift of the conductance dips, via Eq.\ (\ref{eq:linearEalpha}). The energy shift, i.e.\ the change in Fermi energy, is controlled by
the chemical potential of the leads.
The backgate that
controls the spin-orbit strength introduces an extra shift in the Fermi energy (due to the electrostatic coupling of the backgate to the 2DEG).  This shift can be compensated by changing the chemical 
potential of the leads by an equal amount.  Similar methods have been used to probe the energy spectrum of quantum dots using transport 
methods \cite{KouwenhovenRepProgPhys2001}.  Another way to compensate for this
shift is  to use a combination of
back and front gates.  This has  been experimentally demonstrated to control the Rashba spin-orbit interaction strength without introducing
charging in the 2DEG\cite{Grundler:prl2000}.

\section{Variations of parameters in the periodic potential}
\label{sec:PerVar}
The periodic potential plays a crucial role in in the formation and behavior of the energy gaps. We therefore 
devote this section to study the effect that  variations of parameters in the periodic potential has on the conductance.
In what follows we consider variations on the length and strength of the periodic potential (Sec.\ \ref{sec:VarLenStrength})
as well as random fluctuations on the period and strength of each finger gate potential (Sec.\ \ref{sec:Imprecision}).

\subsection{Changing the periodic potential parameters}
\label{sec:VarLenStrength}
\begin{figure}[htp]
\centering
 \includegraphics[width=0.47\textwidth]{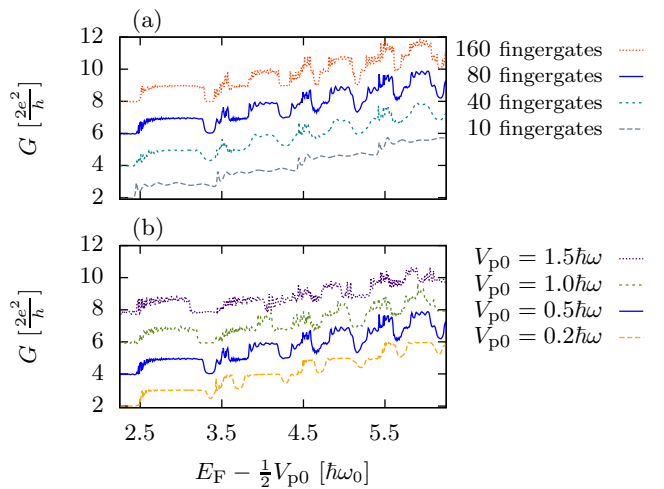}
\caption{(Color online) Charge conductance $G$ through the finite periodic potential as a function of Fermi energy 
$E\Tsub{F}$ for (a) different values of $V\Tsub{p0}$ and (b) different number of fingergates. All the conductances were
calculated with $\alpha=12$ meVnm. Note that the conductances have been separated by $2\times2e^2/\hbar$.}
\label{fig:Diff}
\end{figure}

Here we examine the effects of changing the strength of the periodic potential, $V\Tsub{p0}$, and the
length $L_x$. In Fig.\ \ref{fig:Diff} (a) the length of 
the wire is varied such that it contains 10, 40, 80, or 160 fingergates with a fixed period $\lambda=63$ nm.
There we see the dips easily for $\gtrsim$ 40 fingergates and they become clearer for more fingergates. There is not
much change between the conductance results for 80 and 160 fingergates, i.e. $\approx$ 80 fingergates
is an adequate number of fingergates.
Fig.\ \ref{fig:Diff} (b) shows results where the strength of the periodic potential has been varied between
values of 0.2, 0.5, 1.0, and 1.5 $\hbar\omega_0$. We see that the width of the conductance dips grows with larger $V\Tsub{p0}$.
But as $V\Tsub{p0}$ increases the total conductance strength $G$ deteriorates due to interferences. This 
makes the detection of the conductance dips difficult. 

\subsection{Effects of fluctuations in the periodic potential}
\label{sec:Imprecision}
 \begin{figure}[htp]
  \centering
  \includegraphics[width=0.49\textwidth]{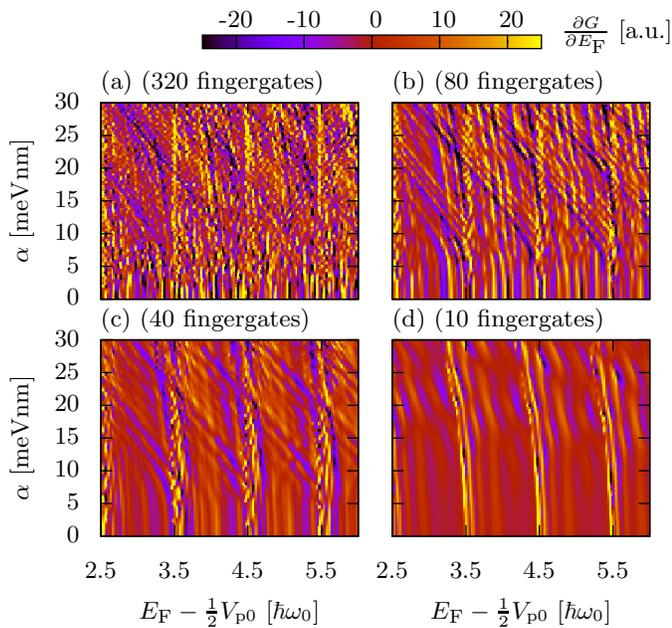}
  \caption{Plots of the differential conductance $\partial G/\partial E\Tsub{F}$ through the finite periodic potential as a function
           of Fermi energy $E\Tsub{F}$ and spin-orbit coupling, $\alpha$. Four cases are considered:
           A system with (a) 320 , (b) 80, (c) 40, and (d) 10 fingergates.
           In all the plots we introduce a 5\% Gaussian error in the length and height of the potential in each period. 
           }
  \label{fig:GaussError}
 \end{figure}

The periodic potential used in the previous section can be considered as being 
ideal. A more realistic case would be if there were some fluctuations in the potential.
To verify the robustness of the results in the previous section we rerun the
simulations with a non-ideal periodic potential. This is done by introducing a $5\%$ Gaussian error
in both the height and length of each potential hill created by the fingergates. 
Figure\ \ref{fig:GaussError} (b) shows that extra noise is added to our conductance result; still the trajectories 
of the conductance dips are fairly visible. To counter against the 
noise introduced by fluctuations we could add extra periods to the periodic potential. This 
helps averaging out the noise, as can be seen in Fig.\ \ref{fig:GaussError} (a) where
we have quadrupled the length of the wire and the number of fingergates. This (self)averaging is
however slow and requires a large number of fingergates. Another solution, as we are mainly 
interested in the slope of the crosspoint trajectories, is actually to reduce the number of fingergates to an optimal number.
As can be seen in Fig.\ \ref{fig:GaussError} (c) we obtain many extra trajectories resulting
from the imprecision of the period. But the slope of the trajectories of the conductance dips
can be easily fitted for $\alpha\gtrsim10$. As expected if we go down to as few as 10 fingergates the 
effects of the periodic potential is nearly washed out.

\section{Conclusion}
We have studied a parabolic quantum wire with Rashba spin-orbit interaction and a longitudinal periodic potential.
We find that the energy gaps resulting from the periodic potential split up and shift from the Bragg plane due to the Rashba 
spin-orbit interaction. Above a certain spin-orbit strength some of the new energy gaps shift linearly in energy as a function of the Rashba spin-orbit strength.
We propose that this effect can be used to measure the change in the Rashba spin-orbit strength, e. g. resulting from a voltage gate\cite{Nitta:prl1997}.
The energy gaps result in dips in the charge conductance. The energy shift of these dips can be fitted, by an analytical equation that we derive, Eq.\ (\ref{eq:linearEalpha}),
to extract the strength of the Rashba spin-orbit interaction. The advantage of the this method is that it only requires conductance measurement and not
any external magnetic or radiation source. 

\appendix
\section{Energy crossing points}
In this appendix we calculate the crossing point $k\Tsub{c}$ of the $\ket{k,m,-1}$ and $\ket{k-2\pi/\lambda,m-1,+1}$ energy bands.
We need to solve the equation
\begin{align}
\varepsilon^0_{m,\downarrow}(k,q\Tsub{R})-\varepsilon^1_{m-1,\uparrow}(k,q\Tsub{R})=\frac{2\pi}{\lambda}k-\frac{2\pi^2}{\lambda^2}+1&\nonumber\\
                                                        +A(k,q\Tsub{R})-\Delta_{m+1}(k,q\Tsub{R})+\Delta_{m}(-k+2\frac{\pi}{\lambda},q\Tsub{R})&=0
\label{eq:EnCrossEq}
\end{align}
for $k$. Here we used Eq.\ (\ref{eq:EffEnUp}) and (\ref{eq:EffEnDown}) and have defined
\begin{equation}
 A(k,q\Tsub{R})=\frac{q\Tsub{R}^2}{2}\left(\frac{1}{1-2q\Tsub{R}k+\frac{4\pi}{\lambda}q\Tsub{R}}-\frac{1}{1+2q\Tsub{R}k}\right).
\label{eq:A}
\end{equation}
Equation \ref{eq:EnCrossEq} is a nonlinear equation which is hard to solve directly. An easier way is to linearize the $A(k)$ and
$\Delta_m(k)$ functions around the point $k=k\Tsub{c}^*$. This approximation introduces little errors and allows us to write the crossing point
as
\begin{align}
&k\Tsub{c}=\frac{\frac{2\pi^2}{\lambda^2}-1- B(m,q\Tsub{R})+C(m,q\Tsub{R})k\Tsub{c}^*}
                {\frac{2\pi}{\lambda}+C(m,q\Tsub{R})}.
\label{eq:kCross}
\end{align}
where
\begin{align}
 B(m,q\Tsub{R})&=A(k\Tsub{c}^*,q\Tsub{R})-\Delta_{m+1}(k\Tsub{c}^*,q\Tsub{R})\nonumber\\
               &+\Delta_{m}(-k\Tsub{c}^*+2\pi/\lambda,q\Tsub{R}),
\label{eq:B}
\end{align}
and
\begin{align}
 C(m,q\Tsub{R})&=\left. \frac{\partial A(k,q\Tsub{R})}{\partial k}\right|_{k=k\Tsub{c}^*}
              -\left. \frac{\partial \Delta_{m+1}(k,q\Tsub{R})}{\partial k}\right|_{k=k\Tsub{c}^*}\nonumber\\
              &-\left. \frac{\partial \Delta_{m}(k,q\Tsub{R})}{\partial k}\right|_{k=-k\Tsub{c}^*+2\pi/\lambda}.
\label{eq:C}
\end{align}

The trajectories of the crossing points across the energy-Rashba spin-orbit coupling surface are then 
$\varepsilon^0_{m,\downarrow}(k\Tsub{c},q\Tsub{R})$ or $\varepsilon^1_{m-1,\uparrow}(k\Tsub{c},q\Tsub{R})$.
The crossing points and their trajectories for the other energy bands, namely the ones corresponding to the states
$\ket{k,m+1,+1}$ and $\ket{k-2\pi/\lambda,m,-1}$, the $\ket{k,m,-1}$ and $\ket{k-2\pi/\lambda,m,-1}$,
and the $\ket{k,m+1,+1}$ and $\ket{k-2\pi/\lambda,m-1,+1}$, can be worked out in the same way.

 The $k$ value of the crossing point between the $\varepsilon^0_{m,\downarrow}(k,q\Tsub{R})$ and $\varepsilon^1_{m-1,\uparrow}(k,q\Tsub{R})$ bands remains almost
 constant as a function of $\alpha$.
 This is because these bands move at similar rate in opposite directions in $k$ space
 as $\alpha$ is changed.  
 We take an advantage of this when finding a linear equation around some point $\alpha_0$ in the straight segments
 of the trajectories. We know that $k\Tsub{c}^*$ is a crossing point for $\alpha=\alpha^*$ so we can, instead of
 using Eq.\ (\ref{eq:kCross}), make the approximation that $k\Tsub{c}\approx k\Tsub{c}^*$ for all $\alpha$. We then linearize 
 $\varepsilon^0_{m,\downarrow}(k\Tsub{c}^*,q\Tsub{R})$ by Taylor expanding around $\alpha_0$.
 Solving for $q\Tsub{R}$ in $\varepsilon^0_{m,\downarrow}(k\Tsub{c}^*,q\Tsub{R})$ and scaling back to $\alpha$ we obtain
  \begin{equation}
  \alpha=\frac{\hbar^2}{m^*}\frac{\hbar}{l}\left[
  q\Tsub{R0}
  -\frac{\varepsilon^0_{m,\downarrow}- F(m)}
        { G(m)}\right],
  \label{eq:APPlinearEalpha}
  \end{equation}
where
\begin{align}
 F(m)&=\frac{1}{2}(k\Tsub{c}^*)^2+m+1-\frac{q\Tsub{R0}^2}{2(1+2q\Tsub{R0}k\Tsub{c}^*)}-\Delta_{m+1}(k\Tsub{c}^*,q\Tsub{R0})\nonumber\\
     &=3.7+m\nonumber\\
     &-\sqrt{6.0\times10^{-5}m^3 - 4.1\times 10^{-2}m^2 + 0.20 m+0.79},
\label{eq:F}
\end{align}
and
\begin{align}
 G(m)&=\frac{k\Tsub{c}^*q\Tsub{R0}^2+q\Tsub{R0}}{4(k\Tsub{c}^*)^2q\Tsub{R0}^2+4k\Tsub{c}^*q\Tsub{R0}+1}
       +\left.\frac{\partial\Delta_{m+1}(k\Tsub{c}^*,q\Tsub{R0})}{\partial q\Tsub{R}}\right|_{q\Tsub{R}=q\Tsub{R0}}\nonumber\\
     &=9.8\times10^{-2}\nonumber\\
     &+\frac{5.1\times10^{-4}m^3-2.7\times10^{-2}m^2+0.82m+4.4}{\sqrt{2.4\times10^{-4}m^3-1.6\times10^{-2}m^2+0.79m+3.1}}.
\label{eq:G}
\end{align}
In the last steps of the Eq.\ (\ref{eq:F}) and Eq.\ (\ref{eq:G}) we have plugged in the values $k\Tsub{c}^*=2.35$ and $q\Tsub{R0}=0.537$, which
correspond to $\alpha^*=8$ meVnm and $\alpha_0=20$ meVnm.
Note that all lengths are scaled in the oscillator length $l=\sqrt{\hbar/m^*\omega_0}$  and all energies in $\hbar\omega_0$.

\acknowledgments{This work was supported by the Icelandic Science and Technology Research
Program for Postgenomic Biomedicine, Nanoscience and Nanotechnology, the Icelandic Research Fund, the Research
Fund of the University of Iceland, the Swiss NSF, the NCCRs Nanoscience and QSIT, CNPq, and FAPESP}


\begin{thebibliography}{30}
\expandafter\ifx\csname natexlab\endcsname\relax\def\natexlab#1{#1}\fi
\expandafter\ifx\csname bibnamefont\endcsname\relax
  \def\bibnamefont#1{#1}\fi
\expandafter\ifx\csname bibfnamefont\endcsname\relax
  \def\bibfnamefont#1{#1}\fi
\expandafter\ifx\csname citenamefont\endcsname\relax
  \def\citenamefont#1{#1}\fi
\expandafter\ifx\csname url\endcsname\relax
  \def\url#1{\texttt{#1}}\fi
\expandafter\ifx\csname urlprefix\endcsname\relax\def\urlprefix{URL }\fi
\providecommand{\bibinfo}[2]{#2}
\providecommand{\eprint}[2][]{\url{#2}}

\bibitem[{\citenamefont{Datta and Das}(1990)}]{Datta:apl1990}
\bibinfo{author}{\bibfnamefont{S.}~\bibnamefont{Datta}} \bibnamefont{and}
  \bibinfo{author}{\bibfnamefont{B.}~\bibnamefont{Das}},
  \bibinfo{journal}{Appl. Phys. Lett.} \textbf{\bibinfo{volume}{56}},
  \bibinfo{pages}{665} (\bibinfo{year}{1990}).

\bibitem[{\citenamefont{D.D.~Awschalom}(2002)}]{Awschalom:Book2002}
\bibinfo{author}{\bibfnamefont{N.~S.} \bibnamefont{D.D.~Awschalom},
  \bibfnamefont{D.~Loss}}, \emph{\bibinfo{title}{Semiconductor Spintronics and
  Quantum Computation}} (\bibinfo{publisher}{Springer}, \bibinfo{year}{2002}).

\bibitem[{\citenamefont{Greiner}(1987)}]{Greiner:Book1987}
\bibinfo{author}{\bibfnamefont{W.}~\bibnamefont{Greiner}},
  \emph{\bibinfo{title}{Relativistic quantum mechanics}}
  (\bibinfo{publisher}{Springer}, \bibinfo{year}{1987}).

\bibitem[{\citenamefont{Winkler}(2010)}]{Winkler:book2010}
\bibinfo{author}{\bibfnamefont{R.}~\bibnamefont{Winkler}},
  \emph{\bibinfo{title}{Spin-orbit Coupling Effects in Two-Dimensional Electron
  and Hole Systems}} (\bibinfo{publisher}{Springer Berlin Heidelberg},
  \bibinfo{year}{2010}).

\bibitem[{\citenamefont{Dresselhaus}(1955)}]{Dresselhaus:pr1955}
\bibinfo{author}{\bibfnamefont{G.}~\bibnamefont{Dresselhaus}},
  \bibinfo{journal}{Phys. Rev.} \textbf{\bibinfo{volume}{100}},
  \bibinfo{pages}{580} (\bibinfo{year}{1955}).

\bibitem[{\citenamefont{Bychkov and Rashba}(1984)}]{Bychkov:jpc1984}
\bibinfo{author}{\bibfnamefont{Y.~A.} \bibnamefont{Bychkov}} \bibnamefont{and}
  \bibinfo{author}{\bibfnamefont{E.~I.} \bibnamefont{Rashba}},
  \bibinfo{journal}{Journal of Physics C: Solid State Physics}
  \textbf{\bibinfo{volume}{17}}, \bibinfo{pages}{6039} (\bibinfo{year}{1984}).

\bibitem[{\citenamefont{Engels et~al.}(1997)\citenamefont{Engels, Lange,
  Sch\"apers, and L\"uth}}]{Engels:prb1997}
\bibinfo{author}{\bibfnamefont{G.}~\bibnamefont{Engels}},
  \bibinfo{author}{\bibfnamefont{J.}~\bibnamefont{Lange}},
  \bibinfo{author}{\bibfnamefont{T.}~\bibnamefont{Sch\"apers}},
  \bibnamefont{and} \bibinfo{author}{\bibfnamefont{H.}~\bibnamefont{L\"uth}},
  \bibinfo{journal}{Phys. Rev. B} \textbf{\bibinfo{volume}{55}},
  \bibinfo{pages}{R1958} (\bibinfo{year}{1997}).

\bibitem[{\citenamefont{Nitta et~al.}(1997)\citenamefont{Nitta, Akazaki,
  Takayanagi, and Enoki}}]{Nitta:prl1997}
\bibinfo{author}{\bibfnamefont{J.}~\bibnamefont{Nitta}},
  \bibinfo{author}{\bibfnamefont{T.}~\bibnamefont{Akazaki}},
  \bibinfo{author}{\bibfnamefont{H.}~\bibnamefont{Takayanagi}},
  \bibnamefont{and} \bibinfo{author}{\bibfnamefont{T.}~\bibnamefont{Enoki}},
  \bibinfo{journal}{Phys. Rev. Lett.} \textbf{\bibinfo{volume}{78}},
  \bibinfo{pages}{1335} (\bibinfo{year}{1997}).

\bibitem[{\citenamefont{Bernardes et~al.}(2007)\citenamefont{Bernardes,
  Schliemann, Lee, Egues, and Loss}}]{Bernarder:prb2007}
\bibinfo{author}{\bibfnamefont{E.}~\bibnamefont{Bernardes}},
  \bibinfo{author}{\bibfnamefont{J.}~\bibnamefont{Schliemann}},
  \bibinfo{author}{\bibfnamefont{M.}~\bibnamefont{Lee}},
  \bibinfo{author}{\bibfnamefont{J.~C.} \bibnamefont{Egues}}, \bibnamefont{and}
  \bibinfo{author}{\bibfnamefont{D.}~\bibnamefont{Loss}},
  \bibinfo{journal}{Phys. Rev. Lett.} \textbf{\bibinfo{volume}{99}},
  \bibinfo{pages}{076603} (\bibinfo{year}{2007}).

\bibitem[{\citenamefont{Calsaverini et~al.}(2008)\citenamefont{Calsaverini,
  Bernardes, Egues, and Loss}}]{Calsaverini:prb2008}
\bibinfo{author}{\bibfnamefont{R.~S.} \bibnamefont{Calsaverini}},
  \bibinfo{author}{\bibfnamefont{E.}~\bibnamefont{Bernardes}},
  \bibinfo{author}{\bibfnamefont{J.~C.} \bibnamefont{Egues}}, \bibnamefont{and}
  \bibinfo{author}{\bibfnamefont{D.}~\bibnamefont{Loss}},
  \bibinfo{journal}{Phys. Rev. B} \textbf{\bibinfo{volume}{78}},
  \bibinfo{pages}{155313} (\bibinfo{year}{2008}).

\bibitem[{\citenamefont{Zawadzki and Pfeffer}(2004)}]{Zawadzki:SST2004}
\bibinfo{author}{\bibfnamefont{W.}~\bibnamefont{Zawadzki}} \bibnamefont{and}
  \bibinfo{author}{\bibfnamefont{P.}~\bibnamefont{Pfeffer}},
  \bibinfo{journal}{Semiconductor Science and Technology}
  \textbf{\bibinfo{volume}{19}}, \bibinfo{pages}{R1} (\bibinfo{year}{2004}).

\bibitem[{\citenamefont{Matsuyama et~al.}(2000)\citenamefont{Matsuyama,
  K\"ursten, Mei\ss{}ner, and Merkt}}]{Matsuyama:prb2000}
\bibinfo{author}{\bibfnamefont{T.}~\bibnamefont{Matsuyama}},
  \bibinfo{author}{\bibfnamefont{R.}~\bibnamefont{K\"ursten}},
  \bibinfo{author}{\bibfnamefont{C.}~\bibnamefont{Mei\ss{}ner}},
  \bibnamefont{and} \bibinfo{author}{\bibfnamefont{U.}~\bibnamefont{Merkt}},
  \bibinfo{journal}{Phys. Rev. B} \textbf{\bibinfo{volume}{61}},
  \bibinfo{pages}{15588} (\bibinfo{year}{2000}).

\bibitem[{\citenamefont{Guzenko et~al.}(2007)\citenamefont{Guzenko, Sch\"apers,
  and Hardtdegen}}]{Guzenko:prb2007}
\bibinfo{author}{\bibfnamefont{V.~A.} \bibnamefont{Guzenko}},
  \bibinfo{author}{\bibfnamefont{T.}~\bibnamefont{Sch\"apers}},
  \bibnamefont{and}
  \bibinfo{author}{\bibfnamefont{H.}~\bibnamefont{Hardtdegen}},
  \bibinfo{journal}{Phys. Rev. B} \textbf{\bibinfo{volume}{76}},
  \bibinfo{pages}{165301} (\bibinfo{year}{2007}).

\bibitem[{\citenamefont{Simmonds et~al.}(2008)\citenamefont{Simmonds, Holmes,
  Beere, and Ritchie}}]{Simmonds:jap2008}
\bibinfo{author}{\bibfnamefont{P.~J.} \bibnamefont{Simmonds}},
  \bibinfo{author}{\bibfnamefont{S.~N.} \bibnamefont{Holmes}},
  \bibinfo{author}{\bibfnamefont{H.~E.} \bibnamefont{Beere}}, \bibnamefont{and}
  \bibinfo{author}{\bibfnamefont{D.~A.} \bibnamefont{Ritchie}},
  \bibinfo{journal}{J. Appl. Phys.} \textbf{\bibinfo{volume}{103}},
  \bibinfo{pages}{124506} (\bibinfo{year}{2008}), ISSN
  \bibinfo{issn}{00218979}.

\bibitem[{\citenamefont{Koga et~al.}(2002)\citenamefont{Koga, Nitta, Akazaki,
  and Takayanagi}}]{Koga:prl2002}
\bibinfo{author}{\bibfnamefont{T.}~\bibnamefont{Koga}},
  \bibinfo{author}{\bibfnamefont{J.}~\bibnamefont{Nitta}},
  \bibinfo{author}{\bibfnamefont{T.}~\bibnamefont{Akazaki}}, \bibnamefont{and}
  \bibinfo{author}{\bibfnamefont{H.}~\bibnamefont{Takayanagi}},
  \bibinfo{journal}{Phys. Rev. Lett.} \textbf{\bibinfo{volume}{89}},
  \bibinfo{pages}{046801} (\bibinfo{year}{2002}).

\bibitem[{\citenamefont{Miller et~al.}(2003)\citenamefont{Miller, Zumb\"uhl,
  Marcus, Lyanda-Geller, Goldhaber-Gordon, Campman, and
  Gossard}}]{Miller:prl2003}
\bibinfo{author}{\bibfnamefont{J.~B.} \bibnamefont{Miller}},
  \bibinfo{author}{\bibfnamefont{D.~M.} \bibnamefont{Zumb\"uhl}},
  \bibinfo{author}{\bibfnamefont{C.~M.} \bibnamefont{Marcus}},
  \bibinfo{author}{\bibfnamefont{Y.~B.} \bibnamefont{Lyanda-Geller}},
  \bibinfo{author}{\bibfnamefont{D.}~\bibnamefont{Goldhaber-Gordon}},
  \bibinfo{author}{\bibfnamefont{K.}~\bibnamefont{Campman}}, \bibnamefont{and}
  \bibinfo{author}{\bibfnamefont{A.~C.} \bibnamefont{Gossard}},
  \bibinfo{journal}{Phys. Rev. Lett.} \textbf{\bibinfo{volume}{90}},
  \bibinfo{pages}{076807} (\bibinfo{year}{2003}).

\bibitem[{\citenamefont{Yu et~al.}(2008)\citenamefont{Yu, Dai, Chu, Poole, and
  Studenikin}}]{Yu:prb2008}
\bibinfo{author}{\bibfnamefont{G.}~\bibnamefont{Yu}},
  \bibinfo{author}{\bibfnamefont{N.}~\bibnamefont{Dai}},
  \bibinfo{author}{\bibfnamefont{J.~H.} \bibnamefont{Chu}},
  \bibinfo{author}{\bibfnamefont{P.~J.} \bibnamefont{Poole}}, \bibnamefont{and}
  \bibinfo{author}{\bibfnamefont{S.~A.} \bibnamefont{Studenikin}},
  \bibinfo{journal}{Phys. Rev. B} \textbf{\bibinfo{volume}{78}},
  \bibinfo{pages}{035304} (\bibinfo{year}{2008}).

\bibitem[{\citenamefont{Kato et~al.}(2004)\citenamefont{Kato, Myers, Gossard,
  and Awschalom}}]{Kato:nature2004}
\bibinfo{author}{\bibfnamefont{Y.}~\bibnamefont{Kato}},
  \bibinfo{author}{\bibfnamefont{R.~C.} \bibnamefont{Myers}},
  \bibinfo{author}{\bibfnamefont{A.~C.} \bibnamefont{Gossard}},
  \bibnamefont{and} \bibinfo{author}{\bibfnamefont{D.~D.}
  \bibnamefont{Awschalom}}, \bibinfo{journal}{Nature}
  \textbf{\bibinfo{volume}{427}}, \bibinfo{pages}{50} (\bibinfo{year}{2004}).

\bibitem[{\citenamefont{Duckheim and Loss}(2006)}]{Duckheim:naturephysics2006}
\bibinfo{author}{\bibfnamefont{M.}~\bibnamefont{Duckheim}} \bibnamefont{and}
  \bibinfo{author}{\bibfnamefont{D.}~\bibnamefont{Loss}},
  \bibinfo{journal}{Nature Physics} \textbf{\bibinfo{volume}{2}},
  \bibinfo{pages}{195} (\bibinfo{year}{2006}).

\bibitem[{\citenamefont{Meier et~al.}(2007)\citenamefont{Meier, Salis,
  Shorubalko, Gini, Schön, and Ensslin}}]{Meier:naturephysics2007}
\bibinfo{author}{\bibfnamefont{L.}~\bibnamefont{Meier}},
  \bibinfo{author}{\bibfnamefont{G.}~\bibnamefont{Salis}},
  \bibinfo{author}{\bibfnamefont{I.}~\bibnamefont{Shorubalko}},
  \bibinfo{author}{\bibfnamefont{E.}~\bibnamefont{Gini}},
  \bibinfo{author}{\bibfnamefont{S.}~\bibnamefont{Sch\"{o}n}}, \bibnamefont{and}
  \bibinfo{author}{\bibfnamefont{K.}~\bibnamefont{Ensslin}},
  \bibinfo{journal}{Nature Physics} \textbf{\bibinfo{volume}{3}},
  \bibinfo{pages}{650} (\bibinfo{year}{2007}).

\bibitem[{\citenamefont{Duckheim et~al.}(2009)\citenamefont{Duckheim, Maslov,
  and Loss}}]{Duckheim:prb2009}
\bibinfo{author}{\bibfnamefont{M.}~\bibnamefont{Duckheim}},
  \bibinfo{author}{\bibfnamefont{D.~L.} \bibnamefont{Maslov}},
  \bibnamefont{and} \bibinfo{author}{\bibfnamefont{D.}~\bibnamefont{Loss}},
  \bibinfo{journal}{Phys. Rev. B} \textbf{\bibinfo{volume}{80}},
  \bibinfo{pages}{235327} (\bibinfo{year}{2009}).

\bibitem[{\citenamefont{Golovach et~al.}(2006)\citenamefont{Golovach, Borhani,
  and Loss}}]{Golovach:prb2006}
\bibinfo{author}{\bibfnamefont{V.~N.} \bibnamefont{Golovach}},
  \bibinfo{author}{\bibfnamefont{M.}~\bibnamefont{Borhani}}, \bibnamefont{and}
  \bibinfo{author}{\bibfnamefont{D.}~\bibnamefont{Loss}},
  \bibinfo{journal}{Phys. Rev. B} \textbf{\bibinfo{volume}{74}},
  \bibinfo{pages}{165319} (\bibinfo{year}{2006}).


\bibitem[{\citenamefont{Eldridge et~al.}(2008)\citenamefont{Eldridge, Leyland,
  Lagoudakis, Karimov, Henini, Taylor, Phillips, and Harley}}]{Eldrige:prb2008}
\bibinfo{author}{\bibfnamefont{P.~S.} \bibnamefont{Eldridge}},
  \bibinfo{author}{\bibfnamefont{W.~J.~H.} \bibnamefont{Leyland}},
  \bibinfo{author}{\bibfnamefont{P.~G.} \bibnamefont{Lagoudakis}},
  \bibinfo{author}{\bibfnamefont{O.~Z.} \bibnamefont{Karimov}},
  \bibinfo{author}{\bibfnamefont{M.}~\bibnamefont{Henini}},
  \bibinfo{author}{\bibfnamefont{D.}~\bibnamefont{Taylor}},
  \bibinfo{author}{\bibfnamefont{R.~T.} \bibnamefont{Phillips}},
  \bibnamefont{and} \bibinfo{author}{\bibfnamefont{R.~T.}
  \bibnamefont{Harley}}, \bibinfo{journal}{Phys. Rev. B}
  \textbf{\bibinfo{volume}{77}}, \bibinfo{pages}{125344}
  (\bibinfo{year}{2008}).

\bibitem[{\citenamefont{Jusserand et~al.}(1992)\citenamefont{Jusserand,
  Richards, Peric, and Etienne}}]{Jusserand:prl1992}
\bibinfo{author}{\bibfnamefont{B.}~\bibnamefont{Jusserand}},
  \bibinfo{author}{\bibfnamefont{D.}~\bibnamefont{Richards}},
  \bibinfo{author}{\bibfnamefont{H.}~\bibnamefont{Peric}}, \bibnamefont{and}
  \bibinfo{author}{\bibfnamefont{B.}~\bibnamefont{Etienne}},
  \bibinfo{journal}{Phys. Rev. Lett.} \textbf{\bibinfo{volume}{69}},
  \bibinfo{pages}{848} (\bibinfo{year}{1992}).

\bibitem[{\citenamefont{Mani et~al.}(2004)\citenamefont{Mani, Smet, von Klitzing,
  Narayanamurti, Johnson, and Umansky}}]{Mani:prb2004}
\bibinfo{author}{\bibfnamefont{R.~G.} \bibnamefont{Mani}},
  \bibinfo{author}{\bibfnamefont{J.~H.} \bibnamefont{Smet}},
  \bibinfo{author}{\bibfnamefont{K.} \bibnamefont{von Klitzing}},
  \bibinfo{author}{\bibfnamefont{V.}~\bibnamefont{Narayanamurti}},
  \bibinfo{author}{\bibfnamefont{W.~B.} \bibnamefont{Johnson}},
  \bibnamefont{and} \bibinfo{author}{\bibfnamefont{V.}~\bibnamefont{Umansky}},
  \bibinfo{journal}{Phys. Rev. B} \textbf{\bibinfo{volume}{69}},
  \bibinfo{pages}{193304} (\bibinfo{year}{2004}).

\bibitem[{\citenamefont{Erlingsson et~al.}(2010)\citenamefont{Erlingsson,
  Egues, and Loss}}]{Erlingsson:prb2010}
\bibinfo{author}{\bibfnamefont{S.~I.} \bibnamefont{Erlingsson}},
  \bibinfo{author}{\bibfnamefont{J.~C.} \bibnamefont{Egues}}, \bibnamefont{and}
  \bibinfo{author}{\bibfnamefont{D.}~\bibnamefont{Loss}},
  \bibinfo{journal}{Phys. Rev. B} \textbf{\bibinfo{volume}{82}},
  \bibinfo{pages}{155456} (\bibinfo{year}{2010}).

\bibitem[{\citenamefont{Datta}(1995)}]{Datta:book1995}
\bibinfo{author}{\bibfnamefont{S.}~\bibnamefont{Datta}},
  \emph{\bibinfo{title}{Electronic Transport in Mesoscopic Systems}}
  (\bibinfo{publisher}{Cambridge University Press}, \bibinfo{year}{1995}).

\bibitem[{\citenamefont{Ferry and Goodnick}(1997)}]{Ferry:book1997}
\bibinfo{author}{\bibfnamefont{D.~K.} \bibnamefont{Ferry}} \bibnamefont{and}
  \bibinfo{author}{\bibfnamefont{S.~M.} \bibnamefont{Goodnick}},
  \emph{\bibinfo{title}{Transport in Nanostructures}}
  (\bibinfo{publisher}{Cambridge University Press}, \bibinfo{year}{1997}).

\bibitem[{\citenamefont{Pastawski and Medina}(2001)}]{Pastawski:pmf2001}
\bibinfo{author}{\bibfnamefont{H.~M.} \bibnamefont{Pastawski}}
  \bibnamefont{and} \bibinfo{author}{\bibfnamefont{E.}~\bibnamefont{Medina}},
  \bibinfo{journal}{Rev. Mex. Fis.} \textbf{\bibinfo{volume}{47}},
  \bibinfo{issue}{S1}, \bibinfo{pages}{1-23} (\bibinfo{year}{2001}).

\bibitem[{\citenamefont{Sancho et~al.}(1985)\citenamefont{Sancho, Sancho,
  Sancho, and Rubio}}]{LopezSancho:jpf1985}
\bibinfo{author}{\bibfnamefont{M.~P.~L.} \bibnamefont{Sancho}},
  \bibinfo{author}{\bibfnamefont{J.~M.~L.} \bibnamefont{Sancho}},
  \bibinfo{author}{\bibfnamefont{J.~M.~L.} \bibnamefont{Sancho}},
  \bibnamefont{and} \bibinfo{author}{\bibfnamefont{J.}~\bibnamefont{Rubio}},
  \bibinfo{journal}{Journal of Physics F: Metal Physics}
  \textbf{\bibinfo{volume}{15}}, \bibinfo{pages}{851} (\bibinfo{year}{1985}).

\bibitem[{\citenamefont{Fisher and Lee}(1981)}]{Fisher:prb1981}
\bibinfo{author}{\bibfnamefont{D.~S.} \bibnamefont{Fisher}} \bibnamefont{and}
  \bibinfo{author}{\bibfnamefont{P.~A.} \bibnamefont{Lee}},
  \bibinfo{journal}{Phys. Rev. B} \textbf{\bibinfo{volume}{23}},
  \bibinfo{pages}{6851} (\bibinfo{year}{1981}).

\bibitem[{\citenamefont{Mireles}(2001)}]{Mireles:prb2001}
\bibinfo{author}{\bibfnamefont{F.}~\bibnamefont{Mireles}}, \bibnamefont{and}
  \bibinfo{author}{\bibfnamefont{G.} \bibnamefont{Kirczenow}},
  \bibinfo{journal}{Phys. Rev. B} \textbf{\bibinfo{volume}{64}},
  \bibinfo{pages}{024426} (\bibinfo{year}{2001}).

\bibitem[{\citenamefont{Kouwenhoven}(2001)}]{KouwenhovenRepProgPhys2001}
\bibinfo{author}{\bibfnamefont{L.~P.}~\bibnamefont{Kouwenhoven}}, \bibnamefont{and}
  \bibinfo{author}{\bibfnamefont{D.~G.} \bibnamefont{Austing}}, \bibnamefont{and}
  \bibinfo{author}{\bibfnamefont{S.} \bibnamefont{Tarucha}},
  \bibinfo{journal}{Rep.\ Prog.\ Phys.} \textbf{\bibinfo{volume}{64}},
  \bibinfo{pages}{701} (\bibinfo{year}{2001}).

\bibitem[{\citenamefont{Grundler}(2000)}]{Grundler:prl2000}
\bibinfo{author}{\bibfnamefont{D.}~\bibnamefont{Grundler}},
  \bibinfo{journal}{Phys. Rev. Lett.} \textbf{\bibinfo{volume}{84}},
  \bibinfo{pages}{6074} (\bibinfo{year}{2000}).


\end{thebibliography}
\end{document}